\documentclass[rmp,titlepage,hidelinks,10pt,byrevtex,titlepage,secnumarabic,nofootinbib,longbibliography,a4paper,showkeys,preprintnumbers,twocolumn]{revtex4-2}
\usepackage{graphicx}
\usepackage{amsmath}
\usepackage{amssymb}
\usepackage{float}
\usepackage{moreverb}
\usepackage[final,commandnameprefix=always]{changes}

\usepackage{url}
\usepackage{svg}
\usepackage{tabularx,array}
\usepackage{ccicons}

\newcommand{\zdoi}{20731500}

\newcommand{\zenodo}[1]{\doi{10.5281/zenodo.#1}}

\usepackage{xr-hyper}
\usepackage{xcolor}
\usepackage{colortbl}
\usepackage[datesep=,datetimesep=.,timesep=,showisoZ=true,showseconds=false]{datetime2}
\usepackage[pdftex,
pdfauthor={Jose Maria Martin-Olalla and Jorge Mira},
pdftitle={Methodological guidelines for circadian modeling of Daylight Saving Time: application to the United States},
            pdfsubject={DST and latitude, circadian disruption},
            pdfkeywords={DST; summer time; latitude; sleep deprivation; spring transition; Europe; insolation; season;motor vehicle accidents},
            pdfproducer={Latex with hyperref, or other system},
            pdfcreator={pdflatex, or other tool}]{hyperref}

            \newcommand\rorlink[2]{\href{https://ror.org/#2}{#1\,\includegraphics[scale=0.1]{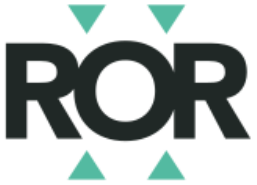}}
}
\usepackage{orcidlink}
\providecommand{\orcidlinki}[2]{\href{https://orcid.org/#2}{#1}\orcidlink{#2}}
            \usepackage{breakurl}
            \usepackage{siunitx}
            \DeclareSIUnit{\uyear}{a}
\usepackage{mathtools}
\usepackage[utf8]{inputenc}
\usepackage[english]{babel}
\usepackage{lineno}
\usepackage{array}

\AtBeginDocument{\usepackage{booktabs}}
\providecommand{\Endparasplit}{}

\graphicspath{{./figures/}{./}}

\DeclareSIUnit{\year}{a}
\makeatletter\AtBeginDocument{\let\@elt\relax}\makeatother

\newcommand{\aucontribute}{\section*{Author contributions}}
\newcommand{\competing}{\section*{Declaration of interest}}
\newcommand{\funding}{\section*{Funding statement}}

\begin{document}
\author{\orcidlinki{José María Martín-Olalla}{0000-0002-3750-9113}}
\affiliation{\rorlink{Universidad de Sevilla}{03yxnpp24}, Facultad de Física, Departamento de Física de la Materia Condensada, ES41012 Sevilla, Spain}
\email{olalla@us.es}

\preprint{\textcolor{blue}{This is a non-peer-reviewed preprint posted on \zenodo{\zdoi}}}
\author{\orcidlinki{Jorge Mira}{0000-0002-6024-6294}}
\affiliation{\rorlink{Universidade de Santiago de Compostela}{030eybx10}, Facultade de Física, Departamento de Física Aplicada and iMATUS, ES15782 Santiago de Compostela, Spain}
\email{jorge.mira@usc.es}
\received[First available: ]{\today}

\thanks{\ccby}
\title{Methodological guidelines for circadian modeling of Daylight Saving Time: application to the United States}
  

\begin{abstract}

  Modeling the circadian impact of seasonal clock changing requires precise synchronization between solar and social time. This report critiques a recent study that associated disease prevalence in the United States with seasonal clock exposure. We identify a fundamental computational error in which a sign reversal of the longitudinal offset effectively inverted the US East-West axis, cross-correlating local health data with the circadian burden of hypothetical locations on the opposite side of a time zone. We outline the methodology for a correct modelization of the circadian process in the context of US geography.

  Word count: 2307

\end{abstract}

\maketitle
The logic underlying seasonal clock adjustments is ostensibly straightforward: it is a synchronized mechanism designed to align morning social schedules with the seasonal oscillations of sunrise triggered by Earth’s axial tilt at extratropical latitudes \cite{Hudson1895}. Yet, this simplicity seems to evaporate when modeling the intricate interplay between immutable solar cycles and plastic human activity.

In this report, we establish a framework for the methodologically robust analysis of circadian models under seasonal time shifts. Our critique is prompted by a recent study attributing disease prevalence across United States counties to seasonal clock exposure \cite{Weed2025}. Our re-analysis reveals fundamental methodological oversights in that work ---most notably, a geographical inversion in which US counties were inadvertently ``flipped.'' This error resulted in cross-correlating local disease prevalence with the hypothetical circadian burden of locations situated on the opposite side of a time zone. Consequently, the original study’s conclusions currently lack empirical support.

Our analysis focuses on the yearly circadian burden. As previously noted, the effects reported in the original study were negligible ---on the order of the model's own time step--- which challenges the validity of associating such minute shifts with bulk societal health outcomes \cite{Martin-Olalla2026c}. Furthermore, the study appears susceptible to ecological fallacy; it cross-correlates individual-level circadian stressors with aggregate societal outcomes, a known source of epidemiological bias \cite{Erren2026}. Our work incorporates a recent correction to the original study.\cite{Weed2026b}

\section{Methods}
\label{sec:methods}

\citet{Weed2025} utilize a circadian model with one regular schedule: 7am-10pm wake, 9am-5pm worktime in weekdays, and free time weekends. Illumination conditions  are fixed to \qty{500}{\lux} at work and to \qty{120}{\lux} in the evening, otherwise determined by ambient light capped at \qty{10}{\kilo\lux}.\cite[Figure~1A]{Weed2025}. Two social clocks are analyzed: the current seasonal clock with two biannual transitions, and a locked clock.

Temporal dynamics are modeled using a 365-day time series with a resolution of $\delta = \qty{5}{\minute}$. In a scientific context, this series represents mean solar time $t_s$, allowing the solar brightness and predefined light conditions to drive the equations for the circadian L-process and P-process uniformly, from which the phase of the circadian clock (the time at core body minimum temperature $\mathrm{CBT}_{\min}$) is obtained.\cite[Figure~2]{Weed2025} Latitude $\phi$ alters solar brightness, sun's height at noon, and sunrise/sunset times with marked seasonality.

The boolean conditions that define light diets ---for instance, between 9:00~am and 5:00~pm Monday to Friday, the illumination is fixed at \emph{worktime} levels---\Endparasplit{} are set by clock time (social time) $t_c$ and must be mapped to solar time to account for the longitudinal gradient created by time zone standardization. The shift is given by the time offset $t_o$, which represents the longitude $\lambda$ offset relative to the corresponding time meridian $\lambda_t$, scaled by Earth's rotation period $\Omega = \qty{15}{\degree\per\hour}$:
\begin{equation}\label{eq:4}t_o = \frac{\lambda - \lambda_t}{\Omega}.\end{equation}
Equation~(\ref{eq:4}) establishes the discrepancy between 12:00 social time and solar noon. West of the time meridian ($\lambda-\lambda_t<0$ and $t_o<0$), solar noon occurs after 12:00 social time. Consequently, $t_s$ lags behind $t_c$: for a fixed $t_c$ (e.g. 9:00~am), the corresponding solar time $t_s=t_c+t_o$ decreases westward as $t_o$ becomes more negative, meaning that $t_s$ is progressively earlier. This behavior is inverse to the conventional observation that sunrises occur later in the West.

Within this framework, Standard Time (ST) and Daylight Saving Time (DST) differ simply by a one-hour shift in $t_o$, as the $\lambda_t$ moves \ang{15} East in the spring and \ang{15} West in the autumn. Consequently, the transition to DST (or to ST) causes $t_o$ to decrease (increase) becoming more (less) negative. The direction of this shift is inverse to the conventional ``spring forward, fall back'' nomenclature.

\begin{figure*}
  \includegraphics{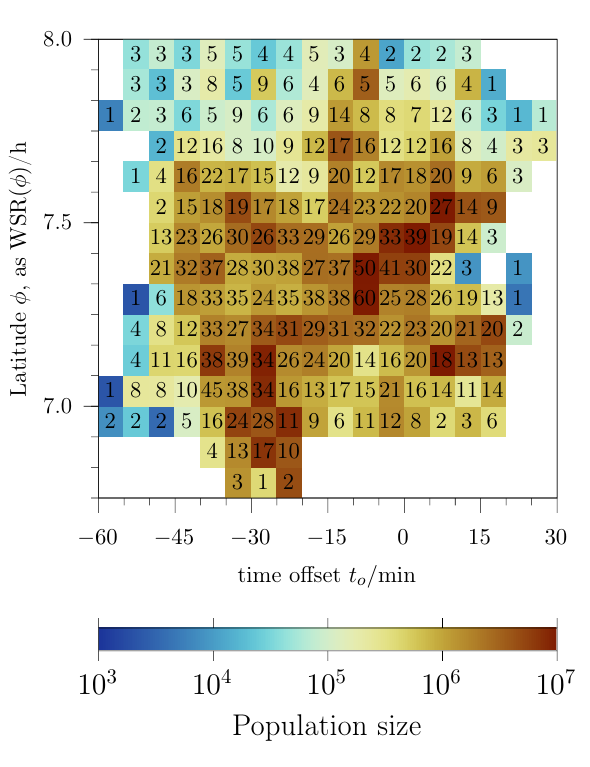}
  \caption{Geographic distribution of US counties. Bin distribution of US counties ($N=\num{3108}$) by centroid time zone offset and latitude, parametrized by winter sunrise time $\mathrm{WSR}(\phi)$. Counties are allocated in \num{207} distinct bins, with \num{29} bins containing at least \num{30} counties.  See figure~\ref{fig:flip} for the distribution of centroids and the values of $\phi$. Palette color is ``roma'', see \citet{Crameri2023}.}
  \label{fig:matrix}
\end{figure*}

Given the model's temporal resolution, $t_o$ can be partitioned into discrete $\qty{5}{\minute}$ bins, ensuring that the underlying boolean conditions remain satisfied within the same time step. Likewise, latitude can be parametrized by the winter sunrise solar time $\mathrm{WSR}$, which acts as a synchronizer of human activity:\cite{Martin-Olalla2018,Martin-Olalla2019b}
\begin{equation}\label{eq:3}\mathrm{WSR}(\phi) = \qty{12}{\hour}-\frac{1}{\Omega} \cos^{-1}(\tan\phi \tan\varepsilon),\end{equation}
where  $\varepsilon = \ang{23.45}$ is Earth's axial tilt. Equation~(\ref{eq:3}) approximates sunrise at the solar center and can be reversed. Consequently, partitioning the spatial domain into \qty{5}{\minute} intervals of $\mathrm{WSR}(\phi)$ is appropriate in this context. Figure~\ref{fig:matrix} illustrates the joint distribution of US counties analyzed in \citet{Weed2025} grouped by their respective $t_o$ and $\mathrm{WSR}(\phi)$ centroid values. Notably, 29 distinct bins contain at least \num{30} counties, serving as a robust data density metric, or a geographic ``h-index'' of sorts, for the analysis of societal outcomes. 

The original study deviate from this approach in two critical aspects. First, the time series was treated as fixed clock time. In line 44 of the script \texttt{computeCircadianShifting.m},\cite{Weed2025b} illumination conditions (\texttt{computeLux}) were corrected by $t_o$ while the boolean conditions that define light diets remained uniform (lines 50 ff.). This reverses physical reality, where solar conditions are uniform at a given latitude while schedules vary, and introduces a false sense of data density that obscures the discrete nature of the time-series and prevents consistency checks.

\begin{figure*}
  \includegraphics[width=\textwidth]{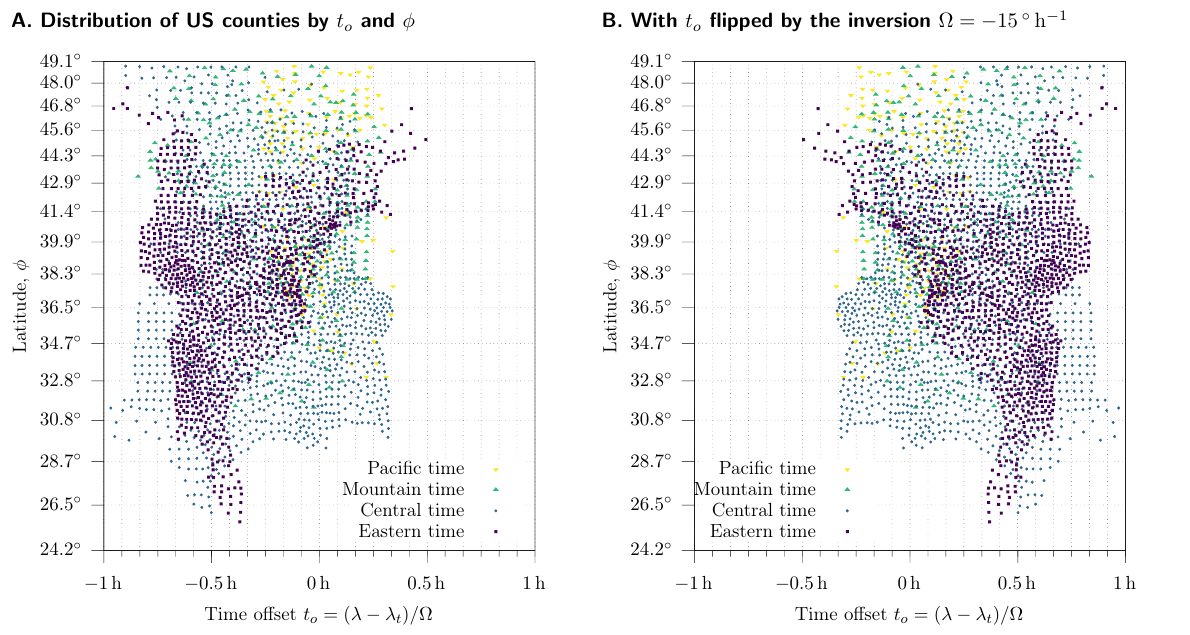}
  \caption{Distribution of US counties by centroid time offset $t_o$ and centroid latitude $\phi$. The grid matches with the bins in figure~\ref{fig:matrix}, which allows to identify bin counts. Distinct US time zones are shown by different shapes and colors. Left: the real distribution with $\Omega=\qty{15}{\degree\per\hour}$. Right: the flipped distribution with $\Omega=\qty{-15}{\degree\per\hour}$ utilized by \citet{Weed2025}.}
  \label{fig:flip}
\end{figure*}

Second, a fundamental sign error occurred when incorporating $t_o$ into lux values. In line 41 of the script, the longitude offset ---named \texttt{LonOffCenter}, computed in line 103 of \texttt{compileCensusData.m}--- is divided by $\Omega=\qty{-15}{\degree\per\hour}$ ---instead of by $\Omega=\qty{15}{\degree\per\hour}$--- to compute $t_o$, a variable named \texttt{eastWest\_hours} in the script. While the script comments correctly state that it is ``brighter earlier in the east'', the instruction to ``so advance the clock'' produced a contradiction: in their model the $t_s$ West of the reference meridian failed to lag behind $t_c$, which is nothing but $t_s$ at the reference meridian. Specifically, at \ang{15} West of the meridian, the brightness belonging to the 10:00~am time step was erroneously assigned to 9:00~am (hence 9:00~am was brighter West of the reference meridian), rather than the 8:00~am brightness required by the longitudinal delay (hence 9:00~am is less bright West of the reference meridian). This error effectively inverted the US East-West axis within the model, see figure~\ref{fig:flip}. Because the original study cross-correlated local disease prevalence of real counties to these geographically ``flipped'' counterpart, the reported associations are based on a mismatch, rendering its conclusions unsupported.

The daily shifting time $\Delta_i$ is the time difference between two consecutive daily phases $\Delta_i=\mathrm{CBT}_{\min}[i+1]-\mathrm{CBT}_{\min}[i]$. The yearly circadian burden $Y$ was computed as $Y=\sum_i|\Delta_i|$.\cite{Weed2025} We observe that by adopting absolute values, this metric is insensitive to the sign of $\Delta_i$ and is consequently mathematically incapable of differentiating between a stable, regulated circadian phase and a non-regulatory phase that monotonically drifts a comparable cumulative value over a year. In contrast, the standard deviation (SD) of the yearly distribution of $\Delta_i$ provides a more physiologically and statistically grounded alternative, specially for the purpose of correlating circadian stressors with health outcomes, the SD represents a more robust metric.

In our framework, we selected $\mathrm{WSR}(\phi)$ values ranging from \qty{6.75}{\hour} ($\phi=\ang{24.2}$) to \qty{8}{\hour} ($\phi=\ang{49}$) while intentionally expanding the analytical scope of $t_o$ from \qty{-3}{\hour} to \qty{0.5}{\hour}. This extended lower bound for $t_o$ was not intended to simulate an extreme ``double Daylight Saving Time'' scenario; rather, it is parametrized to evaluate conditions corresponding to earlier wake and occupational schedules. Within this methodological approach, $t_o$ effectively operationalizes a macro-social time shift relative to the conventional 9:00~am–5:00~pm baseline schedule.

\section{Results and discussion}
\label{sec:results}

\begin{figure*}[t]\centering\includegraphics[width=\textwidth]{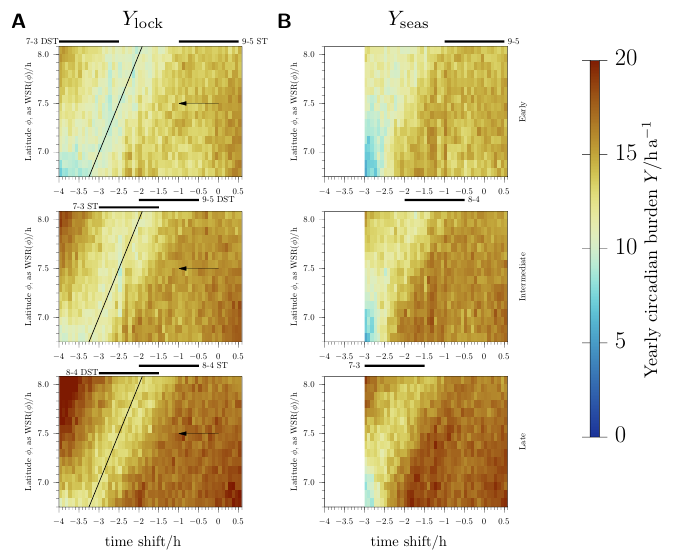}

  \caption{(A) Modeled yearly circadian burden $Y$ under locked, non-seasonal clock policies. (B) $Y$ under the current seasonal alternating clock policy. Color bar on the right indicate the scale range with higher burdens rendered in red, lower burdens in blue. Vertical panels show results for three chronotypes: early ($\tau=\qty{24}{\hour}$, top), intermediate ($\tau=\qty{24.2}{\hour}$, middle), and late ($\tau=\qty{24.4}{\hour}$, bottom). The time shift is relative to the conventional 9:00~am-5:00~pm baseline schedule used to reproduce the model at a given time meridian; moving horizontally twelve bins (\qty{1}{\hour}) to the left is equivalent to moving West by \ang{15} in a time zone, moving the time meridian \ang{15} East (i.e., transitioning to DST); or adopting an early 8:00~am-4:00~pm social schedule. Horizontal bars visualize working hours across the extension of the US (see figure~\ref{fig:matrix}). In panel A the arrows visualize the shift from ST to DST, and the diagonal line (slope equal to 1) shows the coupling with $\mathrm{WSR}(\phi)$.\cite{Martin-Olalla2019b} Palette color is ``roma'', see \citet{Crameri2023}.}
  \label{fig:Y}\end{figure*}

Figure~\ref{fig:Y}A displays a heat map of  $Y$. On the horizontal axis the reference $0$ shows results for the baseline 9:00~am to 5:00~pm at a time meridian. Negative (positive) values represent earlier (later) schedules which can be dynamically operationalized by: (i) moving West (East) within a time zone; (ii) shifting the time meridian East (West) as in transitioning from ST to DST (DST to ST); or (iii) shifting social schedules earlier (later) in absolute time; the horizontal bars between the panels punctuate these shifts, their extension match to figure~\ref{fig:matrix}. The burden generally scales with chronotype, increasing from top to bottom panels. Within a chronotype, a horizontal shift toward the left (see the black arrows) generally decreases the burden within the first third of the plot.

The burden is low: a total of \qty{16}{\hour} per year averages to only \qty{2.63}{\minute} per day, or \qty{0.2}{\percent} of a day. It is improbable that such marginal variations could yield statistically significant societal health outcomes, given the inherent limitations of the model. Furthermore, a theoretical rationale linking the SD or $\sum|\Delta_i|$ ---both metrics of the signal variability--- to health outcomes remains absent.\cite{Weed2025,Martin-Olalla2026c,Weed2026} Figure~\ref{fig:Y}A reveals a dispersed distribution of $Y$ showing perceptible fluctuations for minor changes in $t_o$ and $\mathrm{WSR}$. This dispersion further challenges the validity of $Y$ as a reliable predictor of health risks within the current model.

Generally, transitioning from Standard Time (ST) to Daylight Saving Time (DST) schedules decreases the burden, as illustrated by the horizontal thick bars. A region characterized by a comparatively low $Y$ (rendered in blue) runs diagonally from SW to NE. The overlaid black line features a unit slope, demonstrating a clear coupling between $Y$ (a yearly magnitude) and $\mathrm{WSR}$ (a seasonal quantity) within this region: southerners can sustain earlier schedules without a deterioration in $Y$ as they seldom find dark morning hours.\cite{Martin-Olalla2018,Martin-Olalla2019b} For the specific geography of the US, a conventional social schedule of $t_c = \text{9:00~am}$ corresponds to approximately \qty{1.5}{\hour} after $\mathrm{WSR}$. Consequently, the model suggests that an advance of the social schedule could ameliorate $Y$.

\begin{figure*}
  \includegraphics[width=\textwidth]{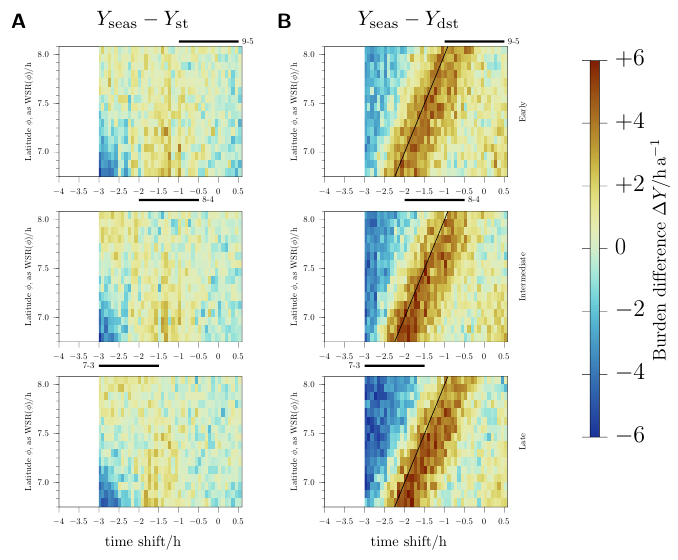}
  \caption{(A) The direct difference $Y_{\text{seas}} - Y_{\text{lock}}$. (B) The direct difference $Y_{\text{seas}} - Y_{\text{lock}}$ with $Y_{\text{lock}}$ shifted \qty{1}{\hour} to the left.  Color bar on the right indicate the scale range with positive differences in red, negative differences in blue. Vertical panels show results for three chronotypes: early ($\tau=\qty{24}{\hour}$, top), intermediate ($\tau=\qty{24.2}{\hour}$, middle), and late ($\tau=\qty{24.4}{\hour}$, bottom). The time shift is relative to the conventional 9:00~am-5:00~pm baseline schedule used to reproduce the model at a given time meridian; moving horizontally twelve bins (\qty{1}{\hour}) to the left is equivalent to moving West by \ang{15} in a time zone, or adopting an early 8:00~am-4:00~pm social schedule. Horizontal bars visualize working hours across the extension of the US (see figure~\ref{fig:matrix}).In panel B the diagonal line (slope equal to 1) shows the coupling with $\mathrm{WSR}(\phi)$.\cite{Martin-Olalla2019b} Palette color is ``roma'', see \citet{Crameri2023}.}
  \label{fig:DY}
\end{figure*}

Figure~\ref{fig:Y}B shows $Y$ for the current US seasonal clock. It largely reproduces the layout of the locked-clock baseline in figure~\ref{fig:Y}A.  Finally While earlier times yield lower $Y$, see the arrows in figure~\ref{fig:Y}A, the biannual clock transitions inflate $Y$ immediately following transition dates by design. These two competing effects essentially cancel out, leaving small net differences, see Figure~\ref{fig:DY}A.

Figure~\ref{fig:DY}B compares the current seasonal clock with the DST clock, obtained by shifting $Y_{\text{lock}}$ \qty{1}{\hour} to the left. Alternatively, a \qty{1}{\hour} displacement left to right in Figure~\ref{fig:Y}A is used. In this scenario the two underlying mechanics reinforce one another: shifting to the right in figure~\ref{fig:Y}A increases $Y$, and the biannual transitions further inflates $Y$, which results in positive $\Delta Y$ in figure~\ref{fig:DY}B. This panel is dominated by a South West to North East band in red (meaning conditions worsened by seasonal clocks) which recalls the diagonal band in figure~\ref{fig:Y}A.

We do not conclude from this specific interaction that permanent DST is superior to permanent ST or the current seasonal clocks. First, individuals aligned with earlier schedules would experience a distinct deterioration of $Y$. Second, because $Y$ is fundamentally low and behaves unsmoothly across narrow parameter windows, it cannot serve as a robust deciding metric for public policy. Notwithstading these limitations, it may provide a subtle quantitative clue regarding the current preferences for DST expressed by polls.\cite{Rubin2023}

The ultimate takeaway of this methodological critique is straightforward: in a scientific context, one must always anchor the analysis to physical reality. Solar conditions must be taken as given; it is the flexible adaptation of social schedules that dictates the alignment. Socially speaking, the current seasonal clock must be understood as an adaptive compromise between two extreme choices.\cite{Martin-Olalla2024g}

\begin{acknowledgments}
  
The authors acknowledge the use of an Artificial Intelligence large language model, Gemini, to refine the draft manuscript for grammar and clarity. Gemini was made available to the authors through a collaborative initiative between Google and Universidad de Sevilla. MatLab\copyright MathWorks 2025b was used to compute the circadian model. MatLab was made available to the authors through a collaborative initiative between MathWorks and Universidad de Sevilla.
\end{acknowledgments}

\aucontribute

All authors contributed equally to this work.

\competing

The authors declare no competing interest.

\funding

This work was not funded.

%apsrmp4-2.bst 2018-12-27 (MD) hand-edited version of apsrmp4-1.bst
%Control: author (3) reversed first dotless

\section*{Manuscript's time line}
After the publishing of our Letter\cite{Martin-Olalla2026c} and reading authors' subsequent reply\cite{Weed2026}, the curiosity bug bit us. We decided to pull down and run the original code ourselves to compute and observe the yearly burden.

The initial execution for the \num{3108} counties took several days, which gave us ample time to dive deep into the underlying methodology. That's when we noticed two critical issues. First, the methodology prioritized clock time over solar time, which is often a really bad take. The issue was hidding at plain sight, but we must admit that when writing our initial Letter we had trusted Authors' framing and did not scrutinize their methodology thoroughly.

Second, as we looked closer, all of a sudden, we uncovered the inversion error. It, too, was sitting right there in plain sight, but we had initially trusted the comments embedded withtin the script rather than checking the raw mechanics. We checked that the light diets were inverted: Western locations had brighter mornings and darker evenings in their clock time analysis.

 The writing of the manuscript required \texttt{28
} different \texttt{emacs} sessions from \texttt{2026-06-17 17:40
} to \mbox{\texttt
{2026-06-17 21:08
}.}

\end{document}